\input epsf.tex

\documentstyle[twocolumn,prb,aps,floats]{revtex}

\def\sumslashD{\mathop{\sum \kern-1.4em -\kern 0.5em}}
\def\sumslash{\mathop{\sum \kern-1.2em -\kern 0.5em}}

\begin{document}
\draft

\title{Interplay between spin-density-wave and superconducting states in quasi-one-dimensional conductors}

\author{ Raphael Duprat and   C. Bourbonnais}

\address{Centre de Recherche sur les Propri\'et\'es \'Electroniques de
Mat\'eriaux Avanc\'es,
}
\address{D\'epartement de Physique, Universit\'e de Sherbrooke, Sherbrooke,
Qu\'ebec, Canada J1K 2R1}

\maketitle
\begin{abstract}
{The interference between spin-density-wave and superconducting instabilities in quasi-one-dimensional
correlated metals is analyzed using the renormalization group method. At the one-loop level,
  we show how the interference leads to a continuous 
crossover from a spin-density-wave state
to unconventional superconductivity when  deviations from perfect nesting of the
Fermi surface exceed a critical value.  Singlet pairing between electrons on neighboring stacks is
found to be the  most favorable symmetry for superconductivity. The consequences of non uniform
spin-density-wave pairing on the structure of phase diagram within the crossover
region is also discussed.
  }
\end{abstract}



\section{Introduction}

 The problem raised by the  interdependence of antiferromagnetism and
superconductivity in low dimensional electronic materials  stands among  the most  important challenges
facing condensed matter physics in  the last two decades or so. Although this issue takes on  considerable
importance in the description  of high-temperature cuprate
superconductors,\cite{Scalapino95,Anderson87,Zhang97} it likely acquired its first focus of interest  in the
context of quasi-one-dimensional organic superconductors, the Bechgaard salts [(TMTSF)$_2$X] and their
sulfur analogs, the Fabre salts [(TMTTF)$_2$X]. The close proximity of antiferromagnetic correlations  
 with the onset of organic superconductivity  
in the temperature and pressure phase diagram of these compounds soon indicated  that the apparent difficulty to
describe both phenomena could find its origin in their mutual
interaction.\cite{Emery86,Bealmonod86,Caron86}

Given the dominant part played by Coulomb repulsion on  the scene
of interactions  in these
materials,\cite{Emery82} 
   early attempts to consider  the nature of superconducting pairing suggested   that   in order to avoid local
repulsion $-$ so resistant to conventional
pairing\cite{Bealmonod86} $-$ electrons  may pair on different
stacks.\cite{Emery83} The
driving force for such a pairing would derive from antiferromagnetic spin
fluctuations,\cite{Emery86,Bealmonod86} a mechanism that can be seen as the
spin counterpart of what Kohn and
Luttinger have proposed long ago for pairing induced by charge (Friedel) oscillations  in the
context of  isotropic
metals.\cite{Kohn65} Its  influence in quasi-one-dimensional metals,
however,  turns out to be more
important than in isotropic materials  extending over a larger domain of temperature in
the normal
phase,\cite{Bourbonnais88,Guay99} and  
  becoming  further enhanced by singular spin-density-wave (SDW) correlations near
the critical pressure $P_c$ above which superconductivity (SC) is singled out as the only stable state.  
An intrinsic difficulty of this problem is that  both SDW  and SC instabilities refer to the same
electron degrees of freedom. Put at the level of  elementary scattering
events  close to the Fermi surface,   electron-hole pairs leading to
density-wave
correlations  interfere with  the electron-electron (hole-hole) pairs
connected with
superconductivity. In 
previous ladder diagrammatic summation,\cite{Emery82,Seidel83,Bealmonod86,Bourbonnais88}
mean-field\cite{Yamaji82,Hasegawa87} and  RPA  
\cite{Shimahara89,Kino99} approaches to  ordered phases at low temperature,  interference is
neglected; an assumption   
 actually grounded on the existence of a coherent warped Fermi surface which is considered as sufficient to
entirely decouple  both types of pairing so that each can be treated separately in   perturbation
theory.\cite{Prigodin79}  However,
as the electron system deconfines at low temperature, namely when a Fermi liquid component can be defined in
at least two spatial directions, interference $-$ of maximum strength in the 1-D non-Fermi
(Luttinger) liquid  domain
$-$ is still present for quasi-particles but  becomes non uniform along the open
Fermi
 surface. It turns out that it is precisely from this uneven
pairing that the interplay between SDW and SC states is found
to  take place.   In practice,
the treatment of both singularities using  the renormalization group (RG) method proved not insurmountable
if interactions within the Fermi liquid  component  are  not too large and a reduction of the number of
variables in flow equations can be made.  Thus at the one-loop RG level, we will show that when
density-wave and superconducting channels interfere on equal footing,\cite{Notes1} the
Kohn-Luttinger mechanism is
dynamically generated from which  a critical threshold for nesting deviations lead to a SDW-SC crossover, a
feature that underlies the possibility of  reentrant superconductivity, in fair agreement with experimental
findings for (TM)$_2$X materials in general.\cite{Jerome82,Jaccard00} As far as  the
nature of superconductivity is concerned,  singlet
interchain pairing is found to be  the  most
favoured symmetry of this ordered state.\cite{Notes2}

 In section II, we introduce the model and define the effective theory for interacting quasi-particles at
low energy. The renormalization group method is  applied in section III and  two-variable  flow equations
for the scattering
 amplitudes are found. The possibilities of long-range order in either density-wave (Peierls)
or superconducting (Cooper) channel are  then analyzed as a function of  deviations from  perfect nesting of the
Fermi surface. The results for the couplings are corroborated by the evaluation of the response functions in
both channels. We close the section III with  considerations on the possibility of reentrant superconductivity
 in the crossover region.  Concluding remarks are given in section IV.

\section{The model at low energy}

We shall base our theoretical description of the competition between antiferromagnetism and
superconductivity on the quasi-one-dimensional electron
gas model. Consider a system
of interacting electrons in a   linear array  of $N_\perp$  chains of
length $L$ ($d_\perp$ is the interchain
distance) and described by the Hamiltonian 
\begin{eqnarray}
&& H=  H_0 + H_I = \sum_{p,{\bf k},\sigma}
E_p({\bf k}) \ a^{\dagger}_{p,{\bf k},\sigma}a_{p,{\bf k},\sigma}
\cr
&& + \ (LN_\perp)^{-1} \pi v_F\sum_{\lbrace {\bf k},{\bf q},\sigma \rbrace}
(g_2\delta_{\sigma_1\sigma_4}\delta_{\sigma_2\sigma_3} -
 g_1\delta_{\sigma_1\sigma_3}\delta_{\sigma_2\sigma_4}) \cr
&& \ \ \ \ \ \ \ \  \ \times \,
a^{\dagger}_{+,{\bf k}_1+{\bf q},\sigma_1}a^{\dagger}_{-,{\bf k}_2-{\bf
q},\sigma_2}
a_{-,{\bf k}_2,\sigma_3}a_{+,{\bf k}_1,\sigma_4}, 
\end{eqnarray}
where
\begin{equation}
E_p({\bf k}) = \epsilon_p (k)-2t_{\perp } \cos(k_{\perp}d_{\perp})
-2t_{\perp2 } \cos(2k_{\perp }d_{\perp}),
\end{equation}
 is the  electron spectrum,  while $\epsilon_p(k)= v_F(pk-k^0_F)$ is the
one-dimensional $-$ linearized $-$ part of the  spectrum  close to the left
and right Fermi points
$pk^0_F=\pm k^0_F$ and  $v_F$ is the 1-D Fermi velocity [${\bf k}\equiv (k,k_\perp)$]. The interchain
single electron hopping  $t_\perp$ is
considered as small compared to the Fermi energy
\hbox{$E_F=v_Fk^0_F$ ($\hbar= 1$)}, which is fixed to be half of the bandwidth
cut-off $E_0\equiv 2E_F$ along
the chains. The even smaller
 transverse hopping term to second nearest-neighbor  chains $t_{\perp2}\ll
t_\perp$
is included in order to parametrize nesting deviations of the entire Fermi
surface. As for the interacting
part
  $H_I$, we  follow the  usual `g-ology'  decomposition  of the direct
interaction between carriers
defined close to the 1-D Fermi points and retain the    backward $g_1$ and  forward
$g_2$ scattering amplitudes $-$ here  normalized by $\pi v_F \ - $ between
right-  and
left-moving carriers.\cite{Solyom79} Given the metallic conditions, one can take into account of  the
influence of umklapp scattering through  a renormalization of  $g_2$ at low energy.
\cite{Emery82,Caron86,Giamarchi97} 

 In the following, we will examine the properties of the model in the gapless regime at
sufficiently low energy scale, that is
when   a coherent interchain motion for  electrons prevails and a Fermi liquid component can be defined.
More specifically, we will
focus on the effective Hamiltonian or $-$ which will be more convenient
for our purposes $-$ on the
effective action of the system  having
$E_x\ll E_0$ as a new bandwidth cut-off for electrons in the neighborhood
of  an open but warped Fermi
surface. This corresponds to the temperature domain $T<E_x/2$ ($k_B=1$)
 of transverse coherence at the single-particle level.
\cite{Bourbon91} 
 Application of the RG  method in the 1-D energy domain
allows the systematic integration of high-energy degrees of freedom.\cite{Bourbon91} The
resulting partition function for our model can then be  expressed as a functional
integral
\begin{eqnarray}
Z= && {\rm Tr }\ e^{-\beta H}\cr
 \sim && \int\!\!\! \int_< D\psi^{\ast}D\psi \ {\rm e}^{S^*\lbrack
\psi^{\ast},\psi \rbrack}
\label{partition}
\end{eqnarray}
over the fermion fields $\psi$ having energies  below $ E_x$. In
the Fourier-Matsubara space, the essential contributions to the effective
action
$S^*$ $-$ up to a renormalization factor for  the fields $-$ can be written
in the form
\begin{eqnarray}
S^*[\psi^*,\psi]= && S_0^*[\psi^*,\psi] + S^*_I[\psi^*,\psi]\cr
                 = && \sum_{p,\sigma} \sum_{\{\tilde{\bf
k}\}^*}\  [G^{0}_{p}(\tilde{\bf k})]^{-1}\
\psi^{\ast}_{p,\sigma}(\tilde{\bf
k})\psi_{p,\sigma}(\tilde{\bf k}) \cr
&& -  \pi v_F\sum_{\mu,\tilde{\bf Q},}
J_\mu(q_\perp,T_{x^1})
O_\mu^*(\tilde{\bf Q})O_\mu(\tilde{\bf Q}) \,
   + \, \ldots,
\label{action}
\end{eqnarray}
where
\begin{equation}
G^0_p(\tilde{\bf k})= \lbrack i\omega_n - E_p^*({\bf k}) \rbrack^{-1},
\end{equation}
is
the `free' propagator in which  the substitutions 
 $t_{\perp(2)} \to
t_{\perp(2)}^* < t_{\perp(2)}$ lead to the renormalization of the   spectrum $ E_p \to
E_p^*$ [here $\tilde{\bf k}= ({\bf
k},\omega_n=\pm\pi T,
\pm 3\pi T,\ldots)$]. The cut-off
 $E_x \approx t_\perp^*$  fixes the maximum energy (or twice the maximum temperature) and the range $\{{\bf
k}\}^*$ of wavevectors in 
the deconfined region. The right (resp.~left) warped
open Fermi surface is parametrized by $k_\perp$
\begin{equation}
\pm k_F(k_\perp) = \pm \big( \, k_F^0 + 2{t^*_\perp\over v_F}\cos(k_\perp
d_\perp) + 2{t_{\perp2}^*\over
v_F}\cos(2k_\perp d_\perp)\, \big).
\end{equation}
 It is
convenient to write the effective interaction between electrons denoted
$S^*_I$ in (\ref{action}) as products of electron-hole pair  fields
\begin{eqnarray}
 O_{\mu}(\tilde{\bf Q}) = && \sqrt{T\over
LN_\perp}\sum_{\alpha,\beta\{\tilde{\bf k}\}^*}
\psi^*_{-,\alpha}(\tilde{\bf k} -\tilde{\bf Q}) \,
\sigma_\mu^{\alpha\beta}\, \psi_{+,\beta}(\tilde{\bf
k})
\label{CompositeP}
\end{eqnarray}
describing  CDW ($\mu=0$) and SDW ($\mu=1,2,3$) correlations, where
$\sigma_{0}$ and $\sigma_{1,2,3}$ are
the identity and Pauli matrices respectively [$\tilde{\bf Q}= (
2k_F^0 +q, q_\perp,\omega_m=0,
\pm 2\pi T,\ldots)$, $q\ll 2k_F^0$]. These correlations are associated to the combinations
of  couplings
\begin{eqnarray}
J_{\mu=0}(q_\perp,\ell_x) = && -{1\over 2} (g^*_2-2g^*_1) + j_{\perp0}\cos
(q_\perp d_\perp),\cr
J_{\mu\ne0}(q_\perp,\ell_x) = && -{1\over 2} g^*_2   + j_{\perp\mu}\cos
(q_\perp d_\perp),
\label{boundary}
\end{eqnarray}
for  the CDW and SDW amplitudes respectively, where the $g_i^*$ are the
renormalized intrachain couplings
evaluated at $\ell_x$ ($E_x=E_0e^{-\ell_x}$). The  $q_\perp$
dependence comes from the interchain
short-range correlations that are generated in the 1-D energy interval from the combined
influence of $g_i$ and $t_\perp$, that is  from $E_0$ down
to $E_x$. The corresponding
interchain pair hopping amplitudes
denoted $j_{\perp\mu}$,  follows from  an explicit RG calculation  and
then modify  the
boundary conditions for the low-energy description.\cite{Bourbon91}  $J_\mu$ is taken  independent
of the longitudinal momentum and Matsubara frequency which is irrelevant (in
the RG sense) in the 1-D domain. This independence is assumed to carry over at lower energy   as implied for
example in the
$k$-dependence
 of boundary conditions (\ref{boundary}), where  the couplings defined with respect to 1-D Fermi points
are used to describe scattering events on   energy edges
$\pm E_x/2$ with respect to a warped Fermi surface.  

\section{Interference between density-wave and superconductivity}
 The nature of correlations that can naturally develop in the deconfined  region for electrons  is
linked in the first place to the amplitude and sign of $J_\mu$. For  repulsive
couplings $g_{1,2}>0$, the  SDW
coupling ($\mu\ne0$) is negative and the
most favorable, while the CDW one, though having the possibility to get the right sign, is
sizably weaker in amplitude at $T>E_x/2$. In the
second place,  the possibility for these  correlations to develop
long-range order at lower temperature is bound to the singular
(logarithmic) response of the system to produce staggered density-wave
correlations; it relies on the
electron-hole  symmetry  of the spectrum, that is on nesting properties. For
an open Fermi
surface, however, the singular response to electron-hole pairing, albeit
weakened by
$t_{\perp2}$, competes with the   response of the Cooper channel for
electron-electron
pairing, which is also singular and unaltered by nesting deviations.      

\subsection{One-loop renormalization group }

The summation of the leading   interfering contributions of the
perturbation theory below
$E_x$ is best  obtained from the application of the  RG
method.  The technique
consists in  partial integrations of $Z$ over fermion degrees of freedom
denoted as $\bar{\psi}^{(*)}$  in the outer energy shell (o.s)
$\pm E_x(\ell)d\ell/2$ above and below the warped Fermi
surface,\cite{Bourbon91} where $E_x(\ell)=E_x
e^{-\ell}$ with
$\ell >0$, is now the scaled bandwidth below $E_x$. Focusing on the
results at the one-loop level, we have
\begin{eqnarray}
Z  &&\ \sim \int\!\!\!\int_< D\psi^*D\psi\, e^{S^*[\psi^*,\psi]_\ell}
\int\!\!\!\int_{o.s} D\bar{\psi}^*D\bar{\psi}\,
e^{S_0^*}\
e^{S^*_{I,2} \, +\, \ldots} \cr
&&\  \propto \int\!\!\!\int_< D\psi^*D\psi\, e^{S^*[\psi^*,\psi]_\ell\  + \
{1\over 2} \langle
S^{*2}_{I,2}
\rangle_{o.s}\  + \ \ldots}
\label{trace}
\end{eqnarray}
where $\langle (S^*_{I,2})^2\rangle_{o.s}$ are the one-loop  outer energy
shell averages calculated with
respect to
$S_0^*[\bar{\psi}^*,\bar{\psi}] $, which  ultimately
lead to the renormalization (flow) of the $J_\mu$ couplings in $S$ as
a function of $\ell$.
Explicitly, by taking $S^*_{I,2}$ as a decomposition of the interaction
having two fields among four to be contracted in the
outer shell and  retaining the contributions coming  from the  Peierls and
Cooper channels, one finds
 \begin{eqnarray}
S^*_{I,2}  = && \,  S^P_{I,2} + S^C_{I,2}\cr
= &&  -\pi v_F\sum_\mu\sumslashD_{\tilde{\bf k}}\sum_{\{\tilde{\bf k}',\tilde{\bf Q}\}^*} 
 J_\mu(q_\perp-k_\perp',k_\perp';\ell)\cr
&&  \ \ \ \ \ \ \ \ \ \ \ \ \ \ \ \ \ \ \ \ \ \times \bar{O}_\mu(\tilde{\bf k}-\tilde{\bf
Q}_0)\,O_\mu(\tilde{\bf k}'-\tilde{\bf Q})\, + {\rm c.c}\cr &&  + \, \pi
v_F\sum_{\bar{\mu}}\sumslashD_{\tilde{\bf k}}\sum_{\{\tilde{\bf k}',\tilde{\bf Q}_c\}^*}
W_{\bar{\mu}}(k_\perp,k'_\perp;\ell)) \cr
&& \ \ \ \ \ \ \ \ \ \ \ \ \ \ \ \ \ \ \ \ \ \times  \,\bar{\cal
O}_{\bar{\mu}}^*(\tilde{\bf k}){\cal
O}_{\bar{\mu}}(\tilde{\bf k}'- \tilde{\bf Q}_c) \, + {\rm c.c}\,,
\label{contraction}
\end{eqnarray}
in which  $\sumslash_{{\bf k}}$ is a sum in the outer momentum shell. 
 Following
(\ref{CompositeP}), we have also defined
$ \sum_{\tilde{\bf k}}O_\mu(\tilde{\bf k}- \tilde{\bf Q}) =
O_\mu(\tilde{\bf Q})$ for the Peierls
fields, whereas  in the Cooper channel,   we have introduced the new fields
\begin{eqnarray}
 {\cal
O}_{\bar{\mu}}(\tilde{\bf k}-\tilde{\bf Q}_c) = && \sqrt{T\over
LN_\perp}\sum_{\alpha,\beta}\alpha\,
\psi_{-,-\alpha}(-\tilde{\bf k} +\tilde{\bf
Q}_c)\sigma_{\bar{\mu}}^{\alpha\beta} \psi_{+,\beta} (\tilde{\bf
k} )
\end{eqnarray}
 for singlet ($\bar{\mu}=0$; SS) and triplet ($\bar{\mu}\ne0$; TS) pairings,
where
$\tilde{\bf Q}_C= (q_C,q_{\perp C},\omega_{mC})
$ corresponds to the momentum and frequency of the Cooper pair. For the fields to be integrated out in 
the contractions (\ref{contraction}), the static Peierls and Cooper external
variables have been set to  $\tilde{\bf
Q}_0=(2k_F^0,q_\perp,0)$ and
$\tilde{\bf Q}_C=0$ respectively.

The contraction in the Cooper channel yields the combinations of
 couplings
\begin{eqnarray}
W_{\bar{\mu}= 0}(k_\perp,k_\perp';\ell) =  &&-{1\over2}
J_0(k_\perp,k_\perp';\ell) \, +\,
{3\over2}J_{\mu\ne0}(k_\perp,k_\perp';\ell)\cr
W_{\bar{\mu}\ne 0}(k_\perp,k_\perp';\ell) =  &&\  {1\over2}
J_0(k_\perp,k_\perp';\ell) \, +\,
{1\over2}J_{\mu\ne0}(k_\perp,k_\perp';\ell)
\label{relations2}
\end{eqnarray}
for SS and TS Cooper pairings.
The explicit evaluation of ${1\over 2}\langle (S^*_{I,2})^2\rangle_{o.s}$,
is given in the
appendix A.
In the case of the Cooper contraction,  one can invert the relations
(\ref{relations}) to re-express
${1\over 2}\langle (S^C_{I,2})^2\rangle_{o,s}
$     in terms of pair fields of the  Peierls
channel. Together with ${1\over 2}\langle (S^P_{I,2})^2\rangle_{o.s}
$ this yields
the   one-loop  flow equations
\begin{eqnarray}
 && {d\over d\ell}J_\mu  (q_\perp-k_\perp,  k_\perp;\ell) = 
{1\over N_\perp}\sum_{\bar{\mu},k_{\perp}'}
c_{\mu,\bar{\mu}} \big[W_{\bar{\mu}}(q_\perp-k_\perp,k_{\perp}';\ell)\cr
&&\ \ \ \ \ \ \ \ \ \ \ \ \ \ \ \ \ \ \ \ \ \ \ \ \ \ \ \ \ \ \ \   \times\   W_{\bar{\mu}}(k_{
\perp}',k_\perp;\ell)\, \big]
I_C(\ell)
\cr
&& -\,J_\mu(q_\perp-k_{\perp},k_{\perp};\ell)
{1\over N_\perp}\sum_{k_{\perp}'}  \big[J_\mu(q_\perp-k_{\perp}',k_{\perp}';\ell) \cr
&&\ \ \ \ \ \ \ \ \ \ \ \ \ \ \ \ \ \ \ \ \ \ \ \ \ \ \ \ \ \ \ \   \times\   I_P(q_\perp,k'_\perp;\ell)\big], 
\label{flow}
\end{eqnarray}
where  the first term  comes  from the Cooper contraction with
the constants $c_{0,0}=-1/2$,
$c_{0,\bar{\mu}\ne0}= 1/2$ and
$c_{\mu\ne 0,0}= 1/2$,
$c_{\mu\ne 0,\bar{\mu}\ne0}=1/6$ and the thermal transient $I_C(\ell)=\tanh
[\beta E_x(\ell)/4]$. As for the Peierls
contraction   in the second term, a kernel
$I_P(q_\perp,k_\perp;\ell)$ follows from the  evaluation of the Peierls
loop which is given  in  appendix A.
Reverting to the initial form  of the action at
$\ell_x$ given in (\ref{action}), we  see  that the interference
between both  channels does not
conserve  the dependence on a  single transverse Peierls variable
$q_\perp$ for  $J_\mu$; the scattering amplitudes now
depends on both ingoing
$[\pm \sim k_F(k_\perp),k_\perp]$  and outgoing $ [\mp \sim k_F(k_\perp
-q_\perp),k_\perp -q_\perp] $ momentum  (note that $J_\mu$ in (\ref{contraction}) is symmetric with respect
to the interchange  $k_\perp
\leftrightarrow k_\perp'$). The  extra
$k_\perp$ dependence leads to  non-uniform  electron-hole pairing, which  gives valuable
information about the strength of the scattering matrix
element and in turn the formation of  the order parameter along the  Fermi
surface. 

 The information  concerning  density-wave or superconducting
instabilities of the normal state
is  obtained from  the singularities  of the  flow equations. Although 2D systems cannot sustain
these forms of long-range order at finite temperature (or finite $\ell$), the temperature $T_i= E(\ell_i)/2$
at which the singularity occurs is nevertheless indicative of the temperature range of true
long-range ordering if a finite coupling in the third direction is added.  Tackling first the possibility of
 a SDW instability in the absence of nesting deviations ($t^*_{\perp2}=0$) and  when   repulsive
interactions
$g_{i=1,2}^*$,
$j_\perp $   prevail at $E_x$, the
 SDW coupling   $J_{\mu\ne0}(q_\perp- k_{\perp},k_\perp; \ell)$ is
found to be  the most
singular  as a function of $\ell$. By way of illustration, if we take $g_1=0.71$, $g_2= 0.80$ and $t_\perp=
160 $K at
$E_F= 3600$K, the 1D two-loop RG results yield, $g_1^*=0.17$, $g^*_2= 0.54$, and
$j_{\perp\mu\ne0(0)}= 0.33(0.024)$   as boundary conditions at \hbox{$E_x= t_\perp^*= 120$}K. Thus
feeding (\ref{flow}) with the latter initial conditions,  the solution of the  flow equation does predict a
singularity at
$
\ell_{SDW}$ corresponding to a critical temperature 	\hbox{$T_{SDW} =
E_x(\ell_{SDW})/2\approx 8.5$}K, which falls in the range of experimental $T_{SDW}$ in (TM)$_2$X when
metallic conditions prevail and nesting deviations ($t_{\perp2}$) are small. \cite{Jerome82} The
singularity  occurs  along the lines
$k_\perp'= \pm\pi/d_\perp  - k_\perp$  (Figure~1a) and  is associated to the SDW
modulation vector
${\bf Q}_P=(2k_F^0,q_\perp=\pi/d_\perp)$ which is the best nesting vector of
the model. On closer examination,  Figure~1a shows that  the scattering amplitude is not uniform, especially in
regions centered at
$k_\perp =\pm
\pi/(2d_\perp)$ (resp. $k_\perp=0,\pm \pi/d_\perp$), which  can be seen as  `cold spots' (resp. hot spots) 
where  a  decrease of $\sim$  50
\% in  the scattering intensity is found. This variation denoted $d(k_\perp)$ in
Figure~\ref{Profile} reflects in turn the one of the SDW order parameter along the Fermi surface.
\cite{Notegap} Therefore the interference with the Cooper channel modifies
 significantly the variation of the SDW gap along the Fermi surface even though nesting deviations are absent.
As we will see below,   these cold regions actually coincide  with the nodes of an interchain superconducting
gap at sufficiently large $t_{\perp2}^*$. 
\begin{figure}
\epsfxsize 8
cm\centerline{\epsfbox{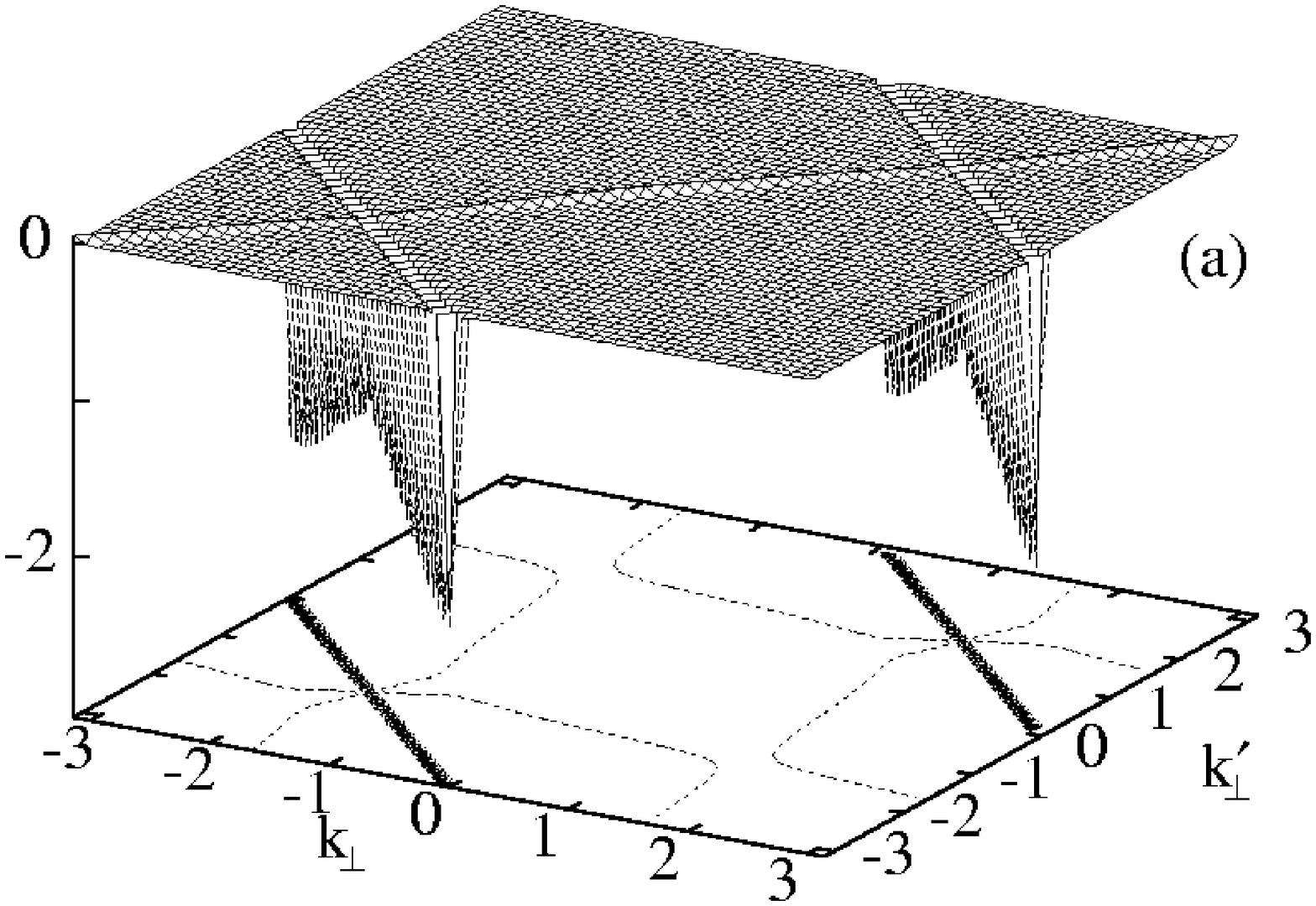}}
\epsfxsize 8
cm\centerline{\epsfbox{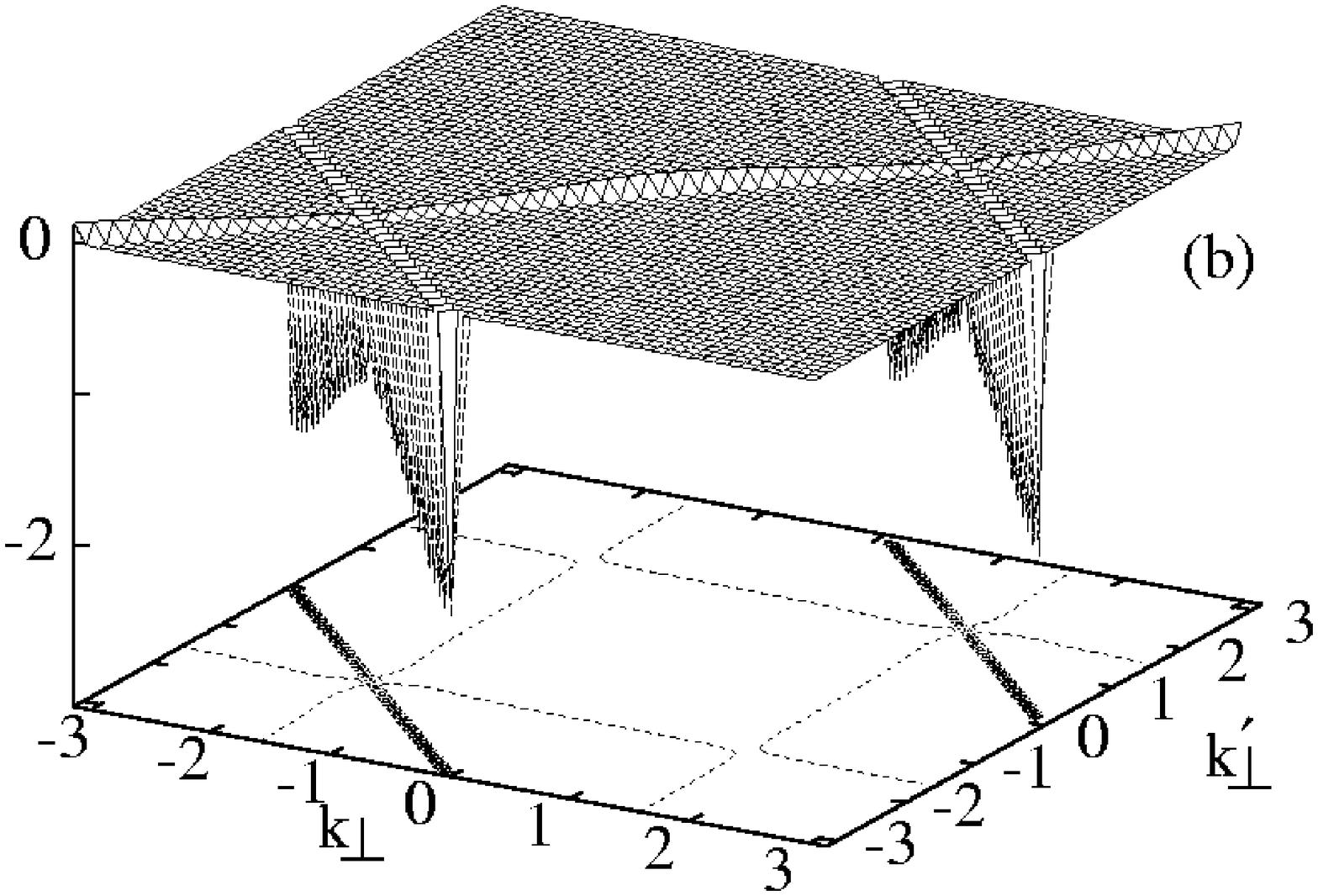}}
\epsfxsize 8
cm\centerline{\epsfbox{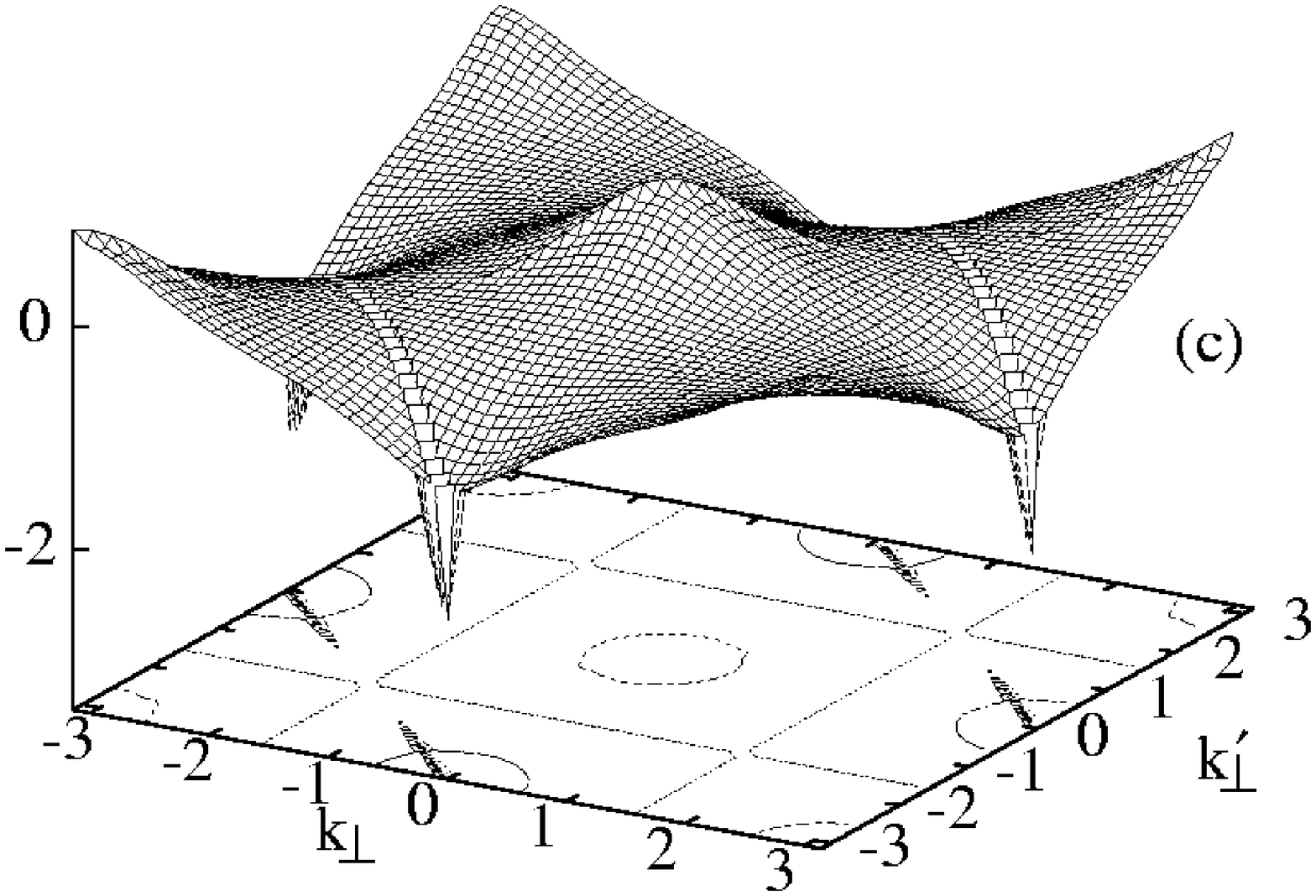}}
\epsfxsize 8
cm\centerline{\epsfbox{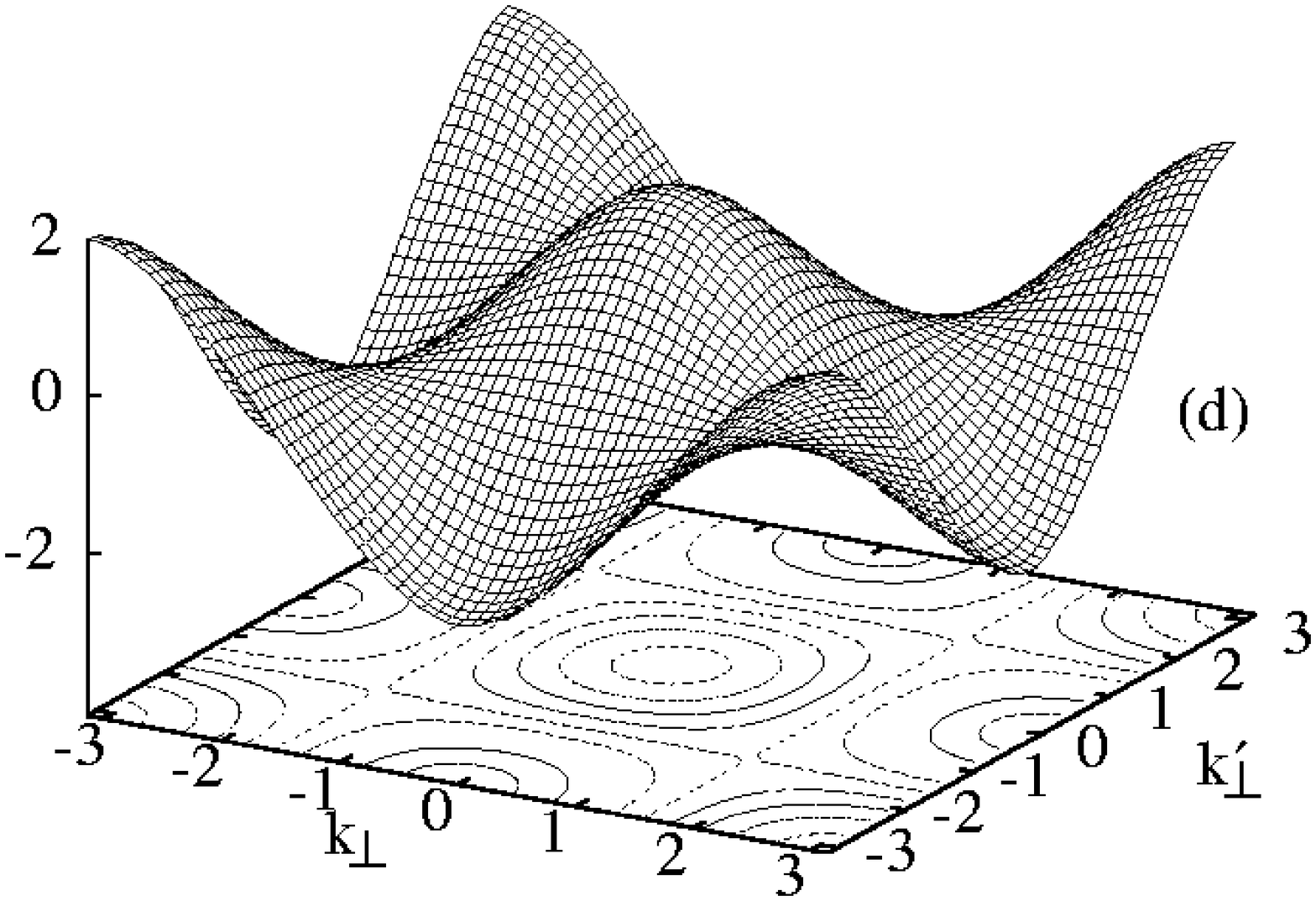}}
\caption{Variation of the SDW coupling $J_{\mu\ne0}(k_\perp,k_\perp';\ell)$ in  the transverse wave vector
$(k_\perp,k_\perp')$ plane close to an instability ($\ell \to \ell_{SDW,c}$). (a):   $t^*_{\perp2}=0$ ;
(b): below  but close to $t^{*c}_{\perp2}= 6.25$K;  (c): crossover region; (d) :   above  the
crossover.  The vertical scale is arbitrary.
  }
\label{Couplings}
\end{figure}
\begin{figure}
\epsfxsize 8cm\centerline{\epsfbox{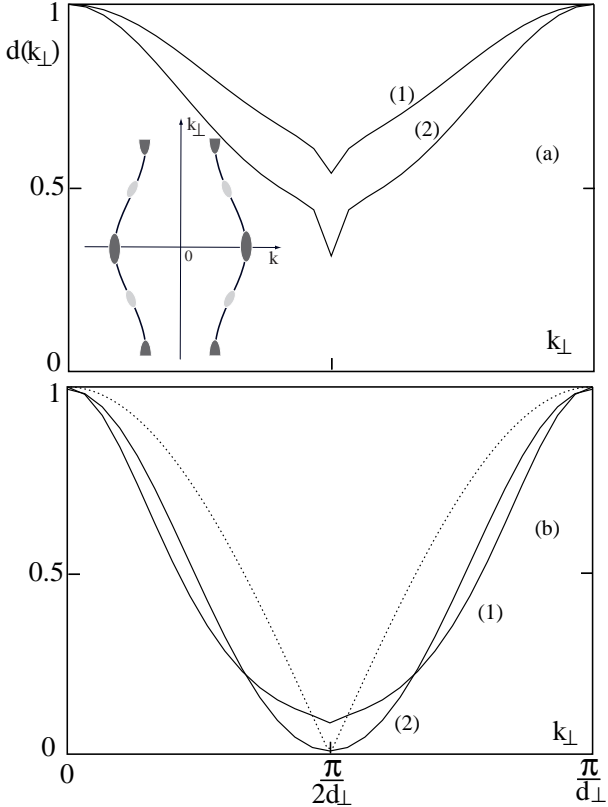}}
\caption{Variation of the SDW normalized scattering amplitude \hbox{$ d(k_\perp)\equiv
J_{\mu\ne0}(\pi/d_\perp- k_\perp, k_\perp)/J^{max}_{\mu\ne0}$} close to
$\ell_{SDW,c}$ or the  gap ($\Delta_{\mu\ne0}\propto
d(k_\perp)$) below $T_{SDW}$  along the Fermi surface $k_F(k_\perp)$ parametrized by $k_\perp$. (a): below
the threshold  for $t^*_{\perp2}=0$ (1),  and $t^*_{\perp2}$ close to $t^{*c}_{\perp2}$ (2); (b): above the
threshold  for
$t^*_{\perp2}$ close to $t^{*c}_{\perp2}$ in the crossover region (1) and $t^*_{\perp2}$ above the
crossover domain (2). The dashed curve  corresponds to the variation of  the interchain singlet gap along
the Fermi surface
\hbox{$|\Delta(k_\perp)|/\Delta_0=|\cos(k_\perp d_\perp)|$}. The inset of (a) shows the location of cold
(light) and hot (dark)  spots on the Fermi surface.  }
\label{Profile}
\end{figure}

\begin{figure}
\vglue 0.4cm\epsfxsize 8cm \centerline{\epsfbox{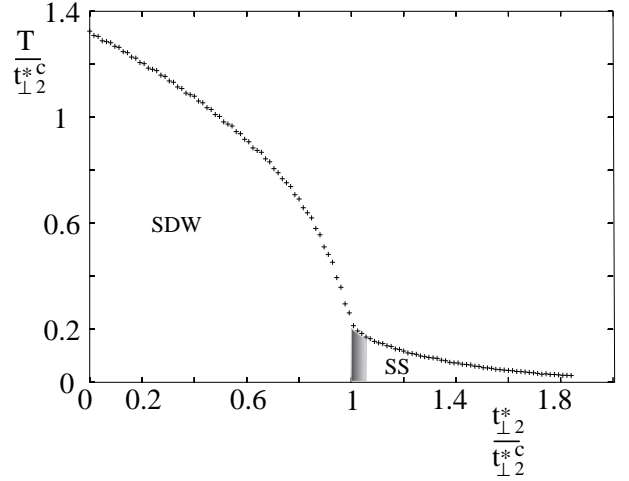}}\vglue 0.4cm
\caption{Variation of the critical temperature as a function of the amplitude of nesting deviations
$t_{\perp2}^*$. The shaded area corresponds to the crossover region where reentrant superconductivity can
occur.}
\label{Tc}
\end{figure}

Now if we try to mimic the influence of  pressure, its effect can be parametrized through the growth  of 
$t^*_{\perp2}$. Thus  by taking  \hbox{$dE_F(v_F)/dt_{\perp2}^*\approx 90$}, the  RG procedure in the 1-D
regime gives \hbox{$d g_{1(2)}^*/dt_{\perp2}^*
\approx - 2.5\%(-4\%)$}/K, 
$dj_{\perp\mu\ne0(0)}/dt_{\perp2}^* \approx -4\%(-8\%)$/K  and $dE_x/dt_{\perp2}^*\approx 5$. These
variations 
 lead to a monotonic decrease of $T_{SDW}$ at small $t_{\perp2}^*$
and
ultimately initiate a rapid
drop of  $T_{SDW}$ (Figure~\ref{Tc}). Along this drop, the electron system, albeit still unstable
to the formation SDW order,  develops a scattering amplitude and in turn an order parameter
that is highly non uniform on
the Fermi surface.  `Hot' (resp.`cold') spots   close to
$(\sim \pm k_F(k_\perp),k_\perp=0,\pm\pi/d_\perp)$ [resp. $(\sim \pm
k_F(k_\perp),k_\perp=\pm\pi/2d_\perp)$] are really
taking shape and their locations on the Fermi surface are kept fixed by
the interaction with Cooper pairing as $t_{\perp2}^*$ varies (Figures~\ref{Couplings},\ref{Profile}). 
This contrasts with the results of mean-field theory of the SDW state for which only nesting deviations are
involved and lead to a qualitatively different   variation of the gap.\cite{Ishiguro90}
As we will see below,  a non uniform gap  leads to  an important reduction of
condensation energy for the SDW state.

  As $t^*_{\perp2}$ is further increased, 
 the critical line  in Fig.~\ref{Tc} develops an inflexion point at $t^{*c}_{\perp2}$
(\hbox{$t^{*c}_{\perp2}\approx 6.25$}K using the above figures) above which  the scattering amplitude
becomes warped by a singular modulation  in the
$k_\perp,k'_\perp$ plane 
 at
$\ell_c$ (Figure~\ref{Couplings}c). This signals an
 instability  at
$T_c=E_x(\ell_c)/2$, which  involves an additional channel connected with  superconductivity. In
order to see  what
type of superconducting  pairing scales to strong coupling, it is natural to make  a
  Fourier decomposition of
$J_\mu $. Given the relation for the scattering amplitude in $S$, that is
\begin{eqnarray}
&& -\sum_\mu \sum_{\{\tilde{\bf k},\tilde{\bf k}',\tilde{\bf Q}\}^*} J_\mu(q_\perp- k_\perp,
k'_\perp;\ell)
O_\mu^*(\tilde{\bf k}+\tilde{\bf Q})\ O_\mu(\tilde{\bf k}'-\tilde{\bf Q})\cr
&&  =  \sum_{\bar{\mu}} \sum_{\{\tilde{\bf k},\tilde{\bf k}',\tilde{\bf Q}_c\}^*}
W_{\bar{\mu}}(k_\perp,
k'_\perp;\ell) \ {\cal O}_{\bar{\mu}}^*(\tilde{\bf k} +\tilde{\bf Q}_c)\ {\cal
O}_{\bar{\mu}}(\tilde{\bf k}'-\tilde{\bf Q}_c)
\end{eqnarray}
and the property $W_{\bar{\mu}}(k_\perp,k'_\perp;\ell) =W_{\bar{\mu}}(-k_\perp,
-k'_\perp;\ell) $, one can write
\begin{eqnarray}
W_{\bar{\mu}}  (k_\perp, && k'_\perp; \ell) =    a_{\bar{\mu}}^0(\ell) \cr \,+
&& \sum^\infty_{m,n>0} \big[
a^{m,n}_{\bar{\mu}}(\ell) 
  \cos(mk_\perp d_\perp)\cos(nk_\perp'd_\perp) \cr
 &&  \ \ \ \ \ \ \ \ \ \ +
b_{\bar{\mu}}^{m,n}(\ell)\sin(mk_\perp d_\perp)\sin(nk_\perp'd_\perp)\bigr],
\end{eqnarray}
which allows to express the interaction in the Cooper channel as a sum of potential
with separate variables
with Fourier coefficients $a^{m,n}_{\bar{\mu}}(\ell)$ and
$b^{m,n}_{\bar{\mu}}(\ell)$ that are scale
dependent (here the coefficients $a^{m,n}$ and $b^{m,n}$ with $m\ne n$ are essentially zero). In this way, to
each Fourier amplitude  corresponds  an {\it interchain} superconducting
coupling of different orbital symmetry in the SS ($\bar{\mu}=0$) and TS
($\bar{\mu}\ne0$) channels. As it is
obvious from the parity of the modulation in Figures~1c-d, the  positive
coefficient 
$a_{\bar{\mu}=0}^{1,1}(\ell)$ of the  first harmonic dominates and  is singular at
$\ell_c$; it corresponds to an instability for
 {\it singlet} ($\bar{\mu}= 0$) pairing between nearest-neighbor chains
having a symmetric orbital  part and
a spin part that is antisymmetric.  From (\ref{relations2}),  a smaller
contribution of the
same Fourier coefficient also applies to  triplet superconductivity
($\bar{\mu}\ne 0$), but this pairing
is not globally antisymmetric and remains inactive
for both short-range and
long-range orders. Actually, our results show that the first positive triplet
pairing coefficient satisfying symmetry
requirements is rather contained in the much smaller  Fourier
coefficient
$b^{2,2}_{\bar{\mu}\ne0}(\ell)$ having odd orbital parity for pairing
between second nearest-neighbor
chains. 

One can then define a  finite $t_{\perp2}^*$ interval above  $t^{*c}_{\perp2}$ in which both SS and SDW
pairings scale to strong   coupling. This region is delimited by a shaded area in Figure~\ref{Tc} and
can be equated with the SDW-SS crossover region. Although a one-loop weak
coupling approach does not allow to make definite conclusions about the actual structure  of the phase
diagram in the crossover domain, from the variation of the  strength of
electron-hole pairing
$J_{\mu\ne0}$ (or the SDW gap) across the Fermi surface (Fig.\ref{Profile}-b), it
is possible   at a more qualitative level to infer   that
superconductivity will be the most stable state at low temperature in that region of the phase diagram (cf.
\S~\ref{reentrant}).

The regular but rapid variation under pressure of $T_c(t_{\perp2}^*)$ above the crossover  in
Fig.~\ref{Tc} can be obviously understood as the
reduction  of the density-wave correlations that feed the Cooper channel.

\subsection{Response functions}
The temperature dependence of the amplitude of correlations  leading to the instabilities of the phase
diagram can be obtained from the calculation of response functions in both Peierls  and Cooper channels.
To do so in the RG framework, we  follow the work of reference \cite{Bourbon91} and couple 
the electrons to a set of source fields   at $E_x$
\begin{eqnarray}     
S_h[\psi^*,\psi] = &&  \sum_{\mu} z_\mu\, \bigl[ O^*_\mu({\bf Q}_P) h_\mu({\bf Q}_P)
\ + \,{\rm c.c}\,\bigr] \cr  + &&   \sum_{\bar{\mu},\{\tilde{\bf k}\}^*} z^{(n)}_{\bar{\mu}}\, \bigl[{\cal
O}^*_{\bar{\mu}}(\tilde{\bf k}) h^{(n)}_{\bar{\mu}}(k_\perp)\, + \,{\rm c.c}\,\bigr],
\end{eqnarray}
where the source fields $h_{\mu}$ are taken independent of the frequency for the static response at
${\bf Q}_P=(2k_F^0,\pi/d_\perp)$  in the Peierls  channel; whereas in the Cooper channel we
are interested in the $n$-th harmonic of the static interstack pairing response at $\tilde{\bf Q}_C=0 $ 
with
$h^{(n)}_{0}(k_\perp)=h^{(n)}_{0}\cos(nk_\perp d_\perp) $ for SS and  $h^{(n)}_{\bar{\mu}\ne
0}(k_\perp)=h^{(n)}_{\bar{\mu}\ne 0}\sin (nk_\perp d_\perp) $ for  TS.  Here the 
$z^{(n)}_{\mu(\bar{\mu})}$ are the corresponding vertex corrections, which for simplicity are put equal to
unity at
$E_x$.\cite{corrections} 

 Performing successive partial traces according to (\ref{trace}), the one-loop corrections  yield  
\begin{eqnarray}
S_h && [\psi^*,\psi]_\ell =  \sum_{\mu} z_\mu(\ell)\bigl[ O^*_\mu({\bf Q}_P) h_\mu({\bf Q}_P)
\ + \,{\rm c.c}\bigr] \cr && \ \ \ \ \ \ \ \ \ +  \sum_{\bar{\mu},\{\tilde{\bf k}\}^*} z^{(n)}_{\bar{\mu}}(\ell)
\bigl[{\cal O}^*_{\bar{\mu}}(\tilde{\bf k}) h^{(n)}_{\bar{\mu}}(k_\perp)\, + \,{\rm c.c}\bigr]\cr 
 &&   \ \ \ \ \ \ \ \ \ + \  \sum_\mu \chi_{\mu}(\ell)\,  h^*_\mu({\bf Q}_P) h_\mu({\bf Q}_P) \, \cr  && \
\ \ \ \ \ \ \ \ +
\  \sum_{\bar{\mu},k_\perp}
\chi^{(n)}_{\bar{\mu}}(\ell)\,  h^{(n)*}_{\bar{\mu}}(k_\perp) h^{(n)}_{\bar{\mu}}(k_\perp),
\label{sources}
\end{eqnarray}
where the pair vertex corrections are obtained from the contractions $\langle S_{I,2}
{S}_{h,2}\rangle_{o.s} $ which involved electron-hole and electron-electron loops whose evaluation is
similar to the one given in Appendix A. The pair vertex parts are then governed by the flow equations
\begin{eqnarray}
\frac{d}{d\ell}\ln z_\mu(\ell)= &&-
{1\over N_\perp}\sum_{k_{\perp}}
\big[J_{\mu}(\pi/d_\perp-k_{\perp},k_{\perp};\ell) \cr
&& \ \ \ \ \ \ \ \ \ \ \  \times\  I_P(\pi/d_\perp,k_{\perp},\ell)\big]
\label{vertexSDW}
\end{eqnarray}
in the Peierls channel and 
\begin{eqnarray}
\frac{d}{d\ell}\ln z^{(n)}_{\bar{\mu}=0}(\ell)&=& {1\over 2} a_{\bar \mu=0}^{n,n}(\ell)
I_C(\ell)\\
\frac{d}{d\ell}\ln z^{(n)}_{\bar{\mu}\ne0}(\ell)&=&{1\over 2}  b_{\bar \mu\ne 0}^{n,n}(\ell)
I_C(\ell),
\label{vertexCooper}
\end{eqnarray}
in the   Cooper channel. 

\begin{figure}
\epsfxsize 7.5cm\centerline{\epsfbox{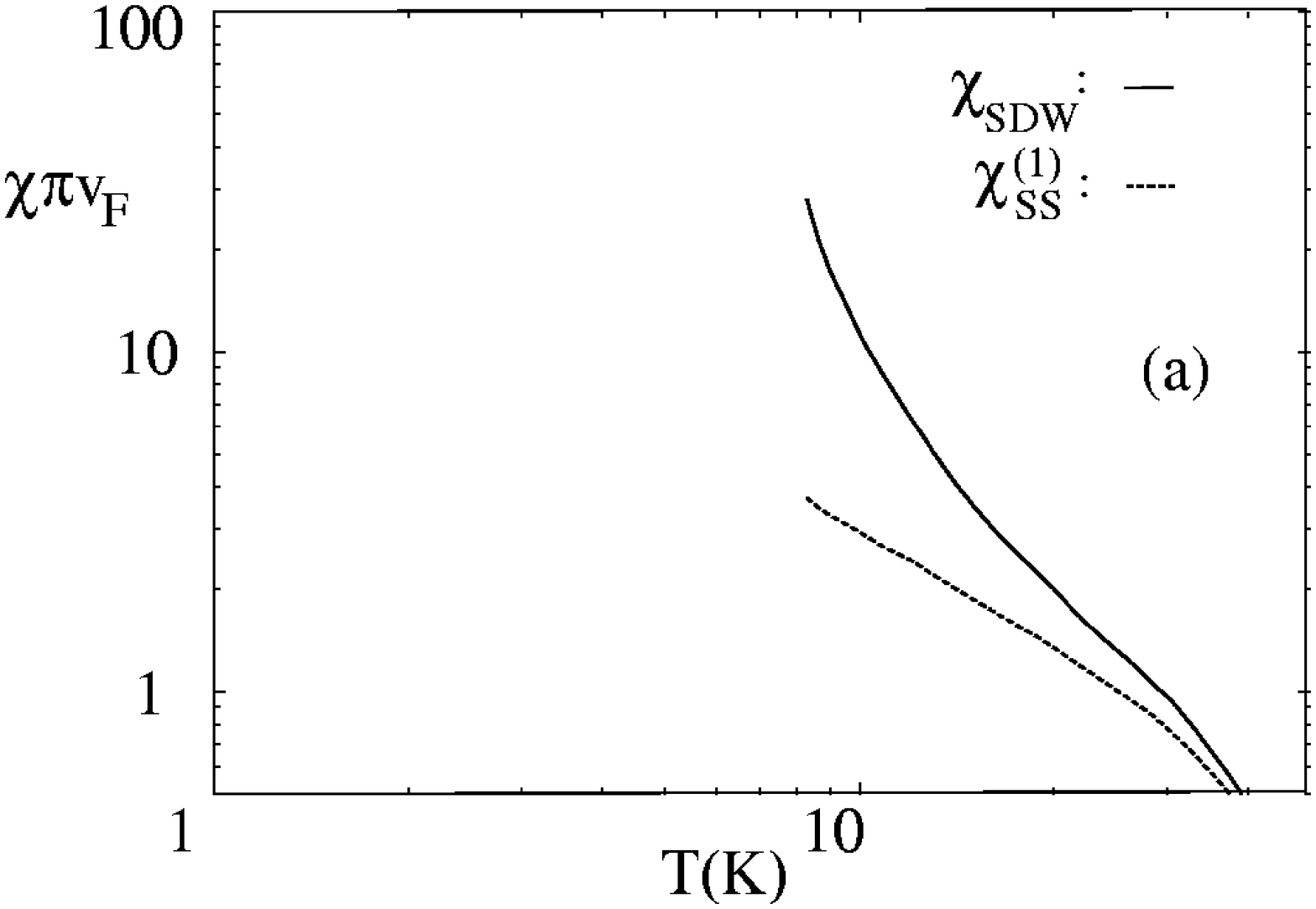}}
\epsfxsize 7.5cm\centerline{\epsfbox{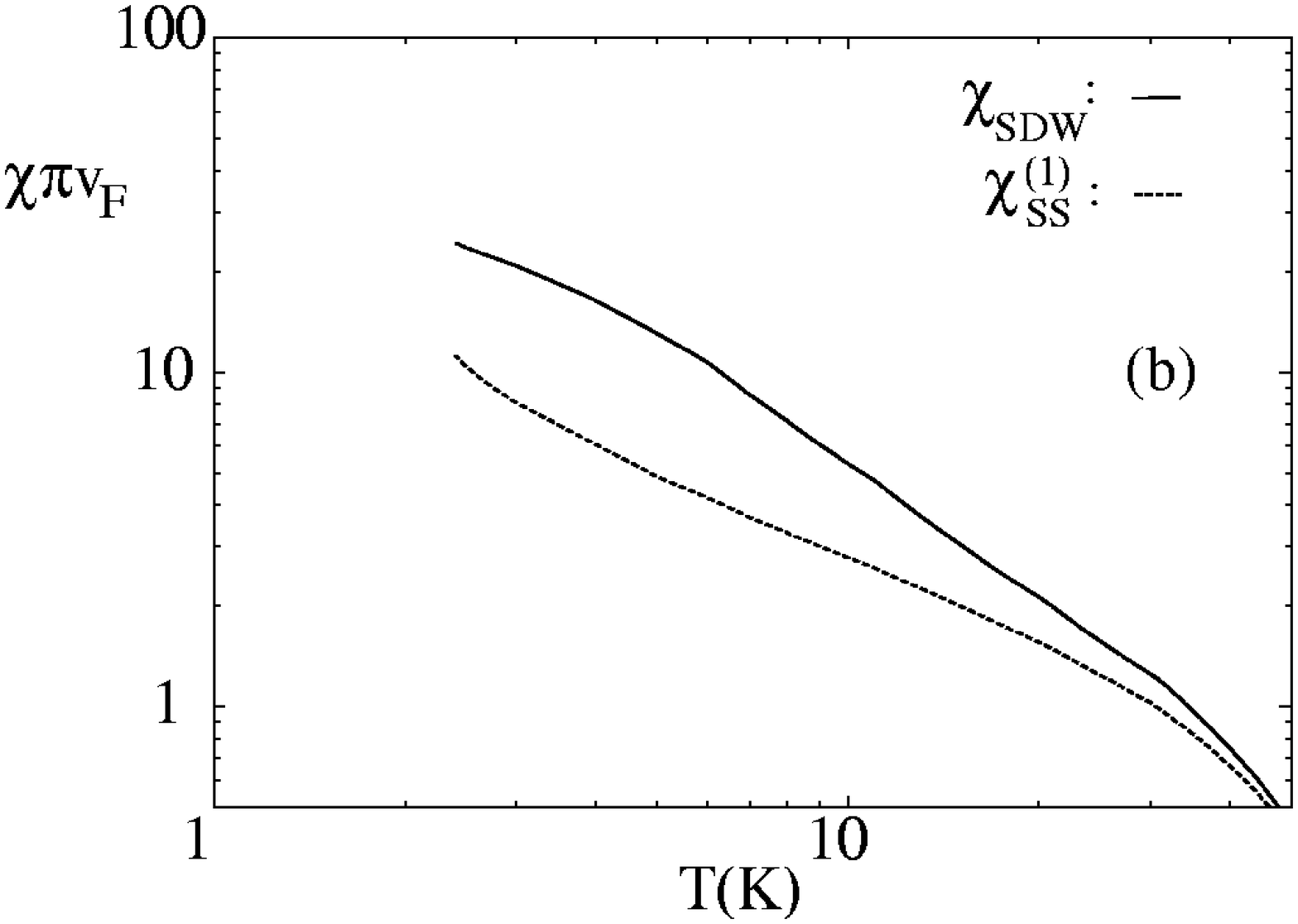}}
\epsfxsize 7.5cm\centerline{\epsfbox{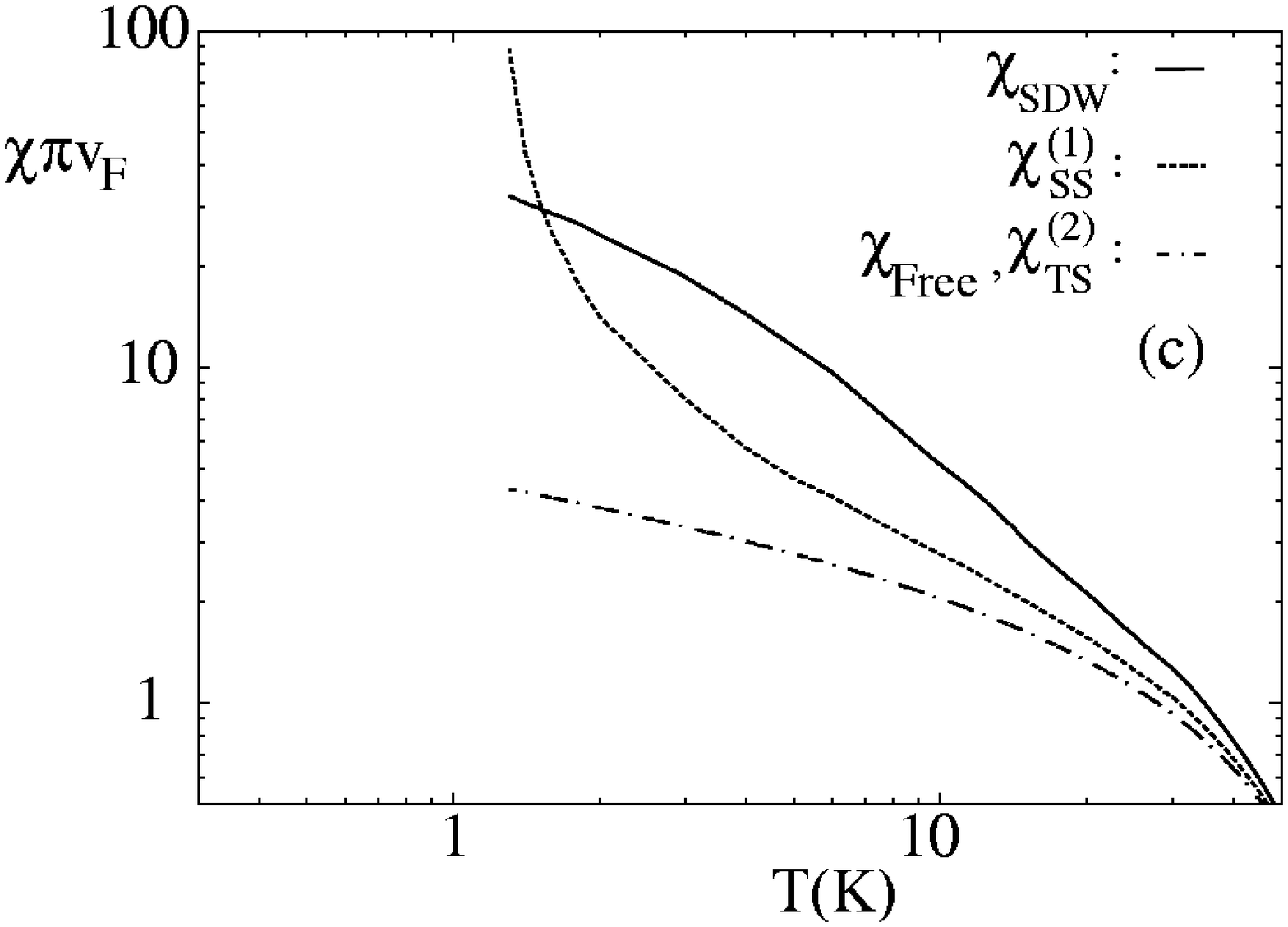}}
\epsfxsize 7.5cm\centerline{\epsfbox{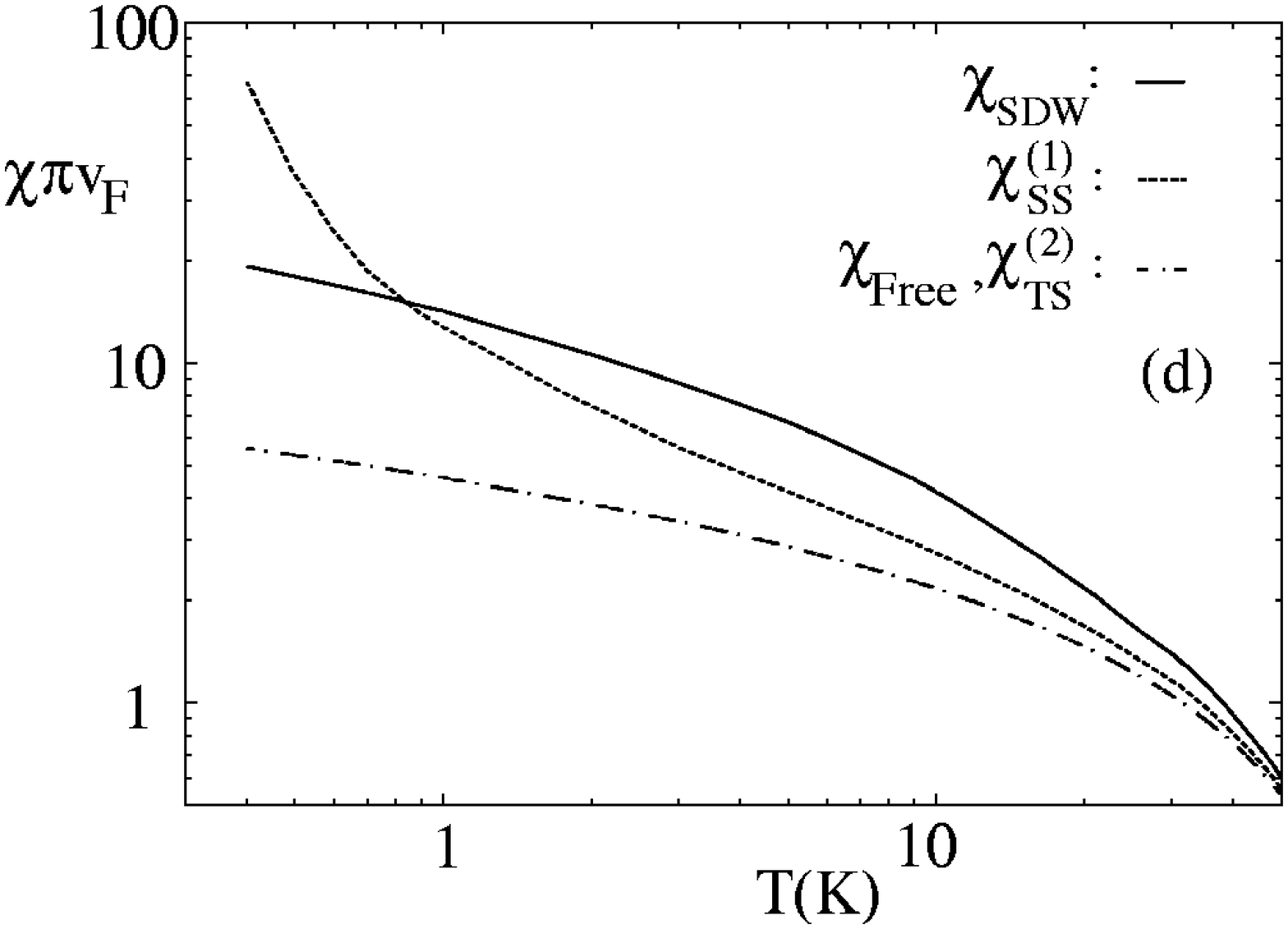}}
\caption{Response functions at $t_{\perp2}^*=0 $ (a), close to but below (b) and above the threshold (c)
$t^{*c}_{\perp2}$, and well above  $t^{*c}_{\perp2}$ (d). }
\label{Responses}
\end{figure}

The second term in (\ref{sources}) comes from the 
contraction
${1\over2}\langle
{S}^2_{h,2}\rangle_{o.s} $, which  generates  expressions that are quadratic in the source fields and whose
coefficients 
\begin{equation}
\chi^{(n)}_{\mu(\bar{\mu})}(\ell) = (\pi v_F)^{-1}\int_0^\ell [z^{(n)}_{\mu(\bar{\mu})}(\ell')]^2 \,
d\ell'
\end{equation}
correspond to the static response function (defined as positive).    From the integration of the flow
equations for the pair vertex parts (\ref{vertexSDW}-\ref{vertexCooper}),  the  temperature dependence  of
SDW and  SC pairing responses are calculated with the set of coupling parameters used in
Figure~\ref{Tc}.  Below
$t^{*c}_{\perp2}$, 
 the SDW response  dominates for  all temperatures and becomes singular at $T_{SDW}$
(Figures~\ref{Responses}-a,b).  In the crossover region, however, both SDW and SS become large but it is
only very close to $T_c$ that the amplitude of SS response catches up with the SDW one
(Figure~\ref{Responses}-c).     Finally above  the crossover (Figure~\ref{Responses}-d), the SS
correlations diverge at $T_c$ whereas SDW correlations, albeit still enhanced, show no sign of singular  
behavior. In Figures~\ref{Responses}-c,d, we verify  that the amplitude of interstack
$n=2$ triplet superconducting  response $\chi^{(2)}_{TS}(T)$  is   scarcely 
enhanced so it essentially merges  with the free or  
non interacting limit $\chi_{Free}(T)$ over the  whole  temperature range.  

\subsection{On the possibility of reentrant  superconductivity}
\label{reentrant}
Before closing this section, we would like to examine on a qualitative basis the consequences of non
uniform pairing on the relative stability of the SDW and SS  order parameters  in the crossover region.
Under the light of the  above results, the profile of the SDW scattering amplitude
$J_\mu(k_\perp-q_\perp,k_\perp;\ell)$, albeit singular on the
lines  $q_\perp=
\pm
\pi/d_\perp$ for all $k_\perp$ as $\ell\to\ell_{SDW,c}$
(Fig.\ref{Couplings}a-c,\ref{Profile},\ref{Responses}a-c),   reflects the variation $d(k_\perp)$
 of the order parameter  and in  turn the
one of the SDW gap
$\Delta_{\mu\ne0}(k_\perp)$ along the Fermi surface within the ordered phase. Therefore if the SDW
transition occurs first, the reduction of the gap in  cold regions will lower the
SDW condensation energy with respect to the case where a uniform gap would be created. This reduction along
with the one of 
$T_{SDW}$ (Fig.~\ref{Tc})   continues up to the crossover region  where the condensation energy becomes
comparable and ultimately weaker than the one of
 a superconducting ordered state having a lower  critical temperature but  a singlet gap
$\Delta(k_\perp)=|\Delta_0|\cos(k_\perp d_\perp)$ 
that would be more developed along the Fermi surface (Figure~\ref{Profile}-b). This crossing of condensation
energy will then signal a first-order transition from SDW  to the SS state $-$ a
possibility  known as reentrant superconductivity
 as pointed out by Yamaji in the framework of a phenomenological mean-field theory of
the competition between SDW and s-wave  states  in the Bechgaard salts.\cite{Yamaji83} 

A quantitative analysis of
reentrance in the present context  would require the microscopic derivation of the temperature dependent
Landau free  energy at $T<T_{SDW}$, which includes interference effects between the
density-wave and Cooper channels, a derivation that  is beyond the scope  of the present paper. More
qualitatively, however, it is possible to estimate the extent of the reentrance  on the $t_{\perp2}^*$
scale by making few assumptions on the form taken by condensation energy in the zero temperature limit. 
Thus if the drop  in energy (here expressed per chain and per unit of length) with respect to the  normal
phase   
 can be expanded in powers of the gap, one can write in lowest order          
\begin{equation}
\delta E_{i}[\Delta_{i}] \simeq -{1\over 2}(\pi v_F)^{-1}C_i\, {1\over N_\perp}\sum_{k_\perp}
\Delta^2_{i}(k_\perp) \, + \, \ldots,
\end{equation}
 for $i=$SDW and SS,  where $C_i$ is a constant (within the mean-field
(BCS) theory, $C_{i}=1$,  and $\delta E_{i}$ is purely quadratic\cite{Mineev99}). Taking
$\Delta_{SDW}(k_\perp)=\, \mid\!\!\Delta_{\mu\ne 0} \!\!\mid d(k_\perp)$ and $\Delta_{SS}(k_\perp)=\,\mid
\!\!\Delta_0\!\!\mid\cos(k_\perp d_\perp)$ for the SDW and a SS gaps respectively and assuming a BCS-like
correspondance
$\mid\!\Delta_i\!\mid\, \sim T_{c,i}$ between the maximum amplitude of the gap at zero temperature  and the
critical temperature, the ratio of condensation energies will take the approximate form
\begin{equation}
{\delta E_{SDW}\over \delta E_{SS}}\approx {T_{SDW}^2\over T_c^2}2 \,N_\perp^{-1}\sum_{k_\perp}
d^2(k_\perp).
\end{equation}
Thus by extracting  $d(k_\perp)$ from the normalized scattering amplitude along $k'_\perp=\pm \pi/d_\perp
-k_\perp$ (Fig.~\ref{Profile}-b),  the  ratio of condensation energies becomes invariably smaller than
unity  for  $T_c$ not too far below $ T_{SDW}$ in the crossover region. This indicates
that whenever  SDW is first stabilized in that part of phase diagram,  the conditions are
favorable for  reentrant superconductivity. The corresponding SDW-SS equilibrium curve should join the
critical line with a positive slope. 
  
\section{Concluding remarks} 
Under the light of previous results interesting conclusions can be drawn. Concerning first the 
  applications to real materials like the Bechgaard salts and their sulfur analogs, the present
approach may serve as a microscopic  basis   towards a synthesis of  itinerant
antiferromagnetism and unconventional superconductivity in  these compounds. This would include 
 the description of systems like the spin-Peierls Fabre salt (TMTTF)$_2$PF$_6$ for which a crossover from
itinerant antiferromagnetism to superconductivity  has been   found  under very high
pressure.\cite{Jaccard00,NotePF6}

The
interplay between SDW and SS correlations has various impacts which may cast new light on both types of
ordering. Starting with the normal phase, the calculation of the response functions shows that both types
of correlations can be enhanced in the same temperature domain, indicating that the different ordering
factors interfere constructively $-$ this is particularly manifest within the narrow confines of reentrant
superconductivity in  the crossover domain.  Another impact  concerns the description of the SDW state
itself, which show qualitative difference with respect to mean-field prediction as  a result of
interference with superconductivity. All goes to show that nesting conditions do not govern  alone  the
characterization of the SDW state, but that Cooper pair correlations have a determining influence leaving
their stamp on  electron-hole pairing or the SDW gap along the Fermi surface. The impact of interference 
on the properties of normal state  in the vicinity of a SDW ordering  in the normal phase is also of interest. It
indeed provides 
  a microscopic basis as to the origin and the location 
 of  cold and hot spots on the Fermi
surface which are suspected to be important in explaining
anomalous magnetotransport effects close to
$t^{*c}_{\perp2}$.\cite{Chaikin92,Zhelezyak95}

The physical picture of  interfering channels in the emergence  of superconductivity naturally bridges to
the physics of the field-induced SDW states for which geometrical properties (nesting) of the Fermi surface
play a so important role in their description.\cite{Gorkov84,Heritier84,Ishiguro90} In this  matter, it
would be worth examining  the weakening of the  interference when a magnetic field is turned on  close to
the crossover on the SS side, namely  where FISDW are found.\cite{Chaikin96}  The presence of
a field (oriented properly)  will weaken both the infrared  singularity of the Cooper channel and   nesting
deviations, which should  tip the balance in favour of a single channel or mean-field 
description.\cite{Gorkov84,Heritier84}  However, it is possible that some traces of interference are still
present  
 and may  affect to some extend the profile of the gap  and then the properties of the FISDW state. 

The present
theoretical weak coupling calculation of the critical temperatures cannot of  course pretend to be
quantitative,  if one considered the reduction of variables in the
obtaining the flow  equations, especially those  related to dynamical (retardation) effects which are 
likely to become relevant within  the crossover region when both SDW and SS pairings scale to strong
coupling. However we think that our approach embodies the essential ingredients of the competition between 
SDW and SC states  in these kind of materials.


\acknowledgements
 C.B thanks  L.G.
Caron, N. Dupuis,  D. J\'erome, D. Jaccard, A.-M. Tremblay and H. Wilhelm for numerous  discussions at
various stages of this work. C.B and R.D also thank  the Natural
Sciences and Engineering Research Council of Canada (NSERC), le Fonds pour
la Formation de Chercheurs et
l'Aide \`a la Recherche du Gouvernement du Qu\'ebec (FCAR), the
`superconductivity program' of the
Institut Canadien de Recherches Avanc\'ees (CIAR)  for financial support.

\appendix
\section{One-loop  contractions}
\subsection{Cooper channel}
 From the last term of the outer shell decomposition in (\ref{contraction}), the
contraction in the Cooper channel evaluated at $\tilde{\bf Q}_c=0$  reads
\begin{eqnarray}
{1\over2}&& \,\langle(S_{I,2}^C)^2\rangle_{o.s} =   (\pi v_F)^2
\sum_{\bar{\mu}_1,\bar{\mu}_2}\sum_{\{\tilde{\bf k},\tilde{\bf k'}
\tilde{\bf Q}_C\}^*}\sumslashD_{\{\tilde{\bf k}_1,\tilde{\bf k}_2\}} \cr && \times \  \langle \,
W_{\bar{\mu}_1}(k_{\perp},k_{\perp,1};\ell)
 \bar{\cal O}^*_{\bar{\mu}_1}(\tilde{\bf
k}_1)\bar{\cal O}_{\bar{\mu}_2}(\tilde{\bf
k}_2)W_{\bar{\mu}_2}(k_{\perp,2},k_{\perp}';\ell)\rangle
\cr
&& \ \ \ \ \ \ \times \ {\cal O}^*_{\bar{\mu}_2}(\tilde{\bf k} +\tilde{\bf Q}_C){\cal
O}_{\bar{\mu}_1}(\tilde{\bf k}'-\tilde{\bf Q}_C) \,
\delta_{\bar{\mu}_1,\bar{\mu}_2}\delta_{\tilde{\bf k}_1,\tilde{\bf k}_2},
\end{eqnarray}
where the outer shell average is given by
\begin{eqnarray}
\sumslashD_{\tilde{\bf k}_1}\ && \langle
W_{\bar{\mu}}(k_{\perp},k_{\perp,};\ell)\bar{\cal
O}^*_{\bar{\mu}}(\tilde{\bf k}_1)  \bar{\cal O}_{\bar{\mu}}(\tilde{\bf k}_1)
W_{\bar{\mu}}(k_{\perp1},k_{\perp}';\ell)\rangle \cr = && 2{T\over
LN_\perp}\sum_{{\omega}_n}\sumslashD_{{\bf
k}_1} \big[ W_{\bar{\mu}}(k_{\perp},k_{\perp1};\ell) G^0_+({\bf k}_1,\omega_n) \cr 
&& \ \ \ \ \ \ \ \ \ \ \ \ \ \ \ \ \ \ \ \times \  G^0_-(-{\bf
k}_1,-\omega_n)W_{\bar{\mu}}(k_{\perp1},k_{\perp}';\ell) \big]\cr
= &&  (\pi v_F)^{-1}{1\over N_\perp}\sum_{k_{\perp1}}
W_{\bar{\mu}}(k_{\perp},k_{\perp1};\ell)W_{\bar{\mu}}(k_{\perp1},k_{\perp}';\ell
) \cr && \ \ \ \ \  \ \ \ \ \ \ \ \ \ \ \ \ \ \ \times \int_{o.s}
 {\tanh{\beta\over 2} E_+\over E_+}\, dE_+,
\end{eqnarray}
where the last line follows from the frequency sum and the use of  the relation 
 $$ 
{d_\perp\over 2\pi}\int\!\!\! \int {dS dE_+\over \mid\nabla E_+\mid}\dots \ = {1\over N_\perp v_F}
\sum_{k_{\perp1}}
\int dE_+\ldots
$$
for the curve integral over constant energy shell. The integral in the outer energy
shell is made over the intervals $[-E_x(\ell)/2, -E_x(\ell + d\ell)/2]$ and
$[E_x(\ell +
d\ell)/2,E_x(\ell)/2]$, and yields the result for the Cooper contraction
\begin{eqnarray}
{1\over2}\, &&\langle(S_{I,2}^C)^2\rangle_{o.s}=  \pi v_F
\sum_{\{\tilde{\bf k},\tilde{\bf k'}
\}^*} {d\ell\over N_\perp}\sum_{k_{\perp1}}
 \big[ W_{\bar{\mu}}(k_{\perp},k_{\perp1};\ell)\cr 
&&   \times \, W_{\bar{\mu}}(k_{\perp1},k_{\perp}';\ell
) \,I_C(\ell)
\, {\cal O}^*_{\bar{\mu}}(\tilde{\bf k}+\tilde{\bf Q}_C){\cal O}_{\bar{\mu}}(\tilde{\bf k}'-\tilde{\bf
Q}_C),
\end{eqnarray}
where $I_C(\ell)= \tanh[\beta E_x(\ell)/4]$. 
Using  the relations 
\begin{eqnarray}
 \sum_{\{\tilde{\bf k},\tilde{\bf k}',\tilde{\bf Q}\}}  O_0^*(\tilde{\bf k}+\tilde{\bf Q})O_0(\tilde{\bf
k}'-\tilde{\bf Q}) =&&  \cr
 \sum_{\{\tilde{\bf k},\tilde{\bf k}',\tilde{\bf Q}_C\}}\big\{ \, {1\over2}{\cal
O}^*_{\bar{\mu}=0}(\tilde{\bf k}+\tilde{\bf Q}_C) {\cal O}_{\bar{\mu}=0}(\tilde{\bf k}'-\tilde{\bf Q}_C)
\, 
 \cr
  - \, {1\over2} \sum_{\bar{\mu}\ne 0} {\cal
O}^*_{\bar{\mu}}(\tilde{\bf k}+\tilde{\bf Q}_C) {\cal O}_{\bar{\mu}}(\tilde{\bf k}'- \tilde{\bf
Q}_C &&)\,\big\}\cr
 \sum_{\{\tilde{\bf k},\tilde{\bf k}',\tilde{\bf Q}\}} \sum_{\mu\ne 0} O_\mu^*(\tilde{\bf k}+\tilde{\bf
Q})O_\mu(\tilde{\bf k}'-\tilde{\bf Q}) = &&  \cr 
 \sum_{\{\tilde{\bf k},\tilde{\bf k}',\tilde{\bf
Q}_c\}} \big\{-{3\over 2}{\cal O}^*_{\bar{\mu}=0}(\tilde{\bf k}+\tilde{\bf Q}_C) {\cal
O}_{\bar{\mu}=0}(\tilde{\bf k}'-\tilde{\bf Q}_C) \,\cr  
\ - \, {1\over2} \sum_{\bar{\mu}\ne 0} {\cal
O}^*_{\bar{\mu}}(\tilde{\bf k}+\tilde{\bf Q}_C) {\cal O}_{\bar{\mu}}(\tilde{\bf k}'-\tilde{\bf
Q}_C && )\,\bigr\}
\label{relations}
\end{eqnarray}
 this result can be
rewritten in terms of products of Peierls fields contributing to the first
term of the flow equation for
$J_\mu$  [Eq. (\ref{flow})].

\smallskip
\subsection{Peierls Channel}
From the first two terms of (\ref{contraction}), the  contractions in the
Peierls channel are given by
\begin{eqnarray}
&& {1\over2}\,\langle(S_{I,2}^P)^2\rangle_{o.s}=  (\pi v_F)^2 \sum_{\mu_1,\mu_2
} \sum_{\{\tilde{\bf
k},\tilde{\bf k'},
\tilde{\bf Q}\}^*}\sumslashD_{\{\tilde{\bf k}_1,\tilde{\bf k}_2\}}\cr && \   \ \times\,  \big[\ \langle \,
J_{{\mu}_1}(q_{\perp}-k_{\perp1},k_{\perp1};\ell)  \bar{
O}^*_{{\mu}_1}(\tilde{\bf
k}_1+\tilde{\bf Q}_0)\bar{ O}_{{\mu}_2}(\tilde{\bf
k}_2 -\tilde{\bf Q}_0)\cr  && \ \ \ \ \ \ \  \times \
J_{{\mu}_2}(q_{\perp}-k_{\perp}',k_{\perp}';\ell)
 { O}^*_{{\mu}_2}(\tilde{\bf k} +  \tilde{\bf Q}){O}_{{\mu}_1}(\tilde{\bf
k}'- \tilde{\bf
Q})
\, \cr  && \ \ \ \ \ \ \ \ \ \ \ \times \delta_{{\mu}_1,{\mu}_2}\delta_{\tilde{\bf k}_1+\tilde{\bf
Q}_0,\tilde{\bf k}_2} \big],
\end{eqnarray}
in which we define 
\begin{eqnarray}
F(q_\perp && )= \sumslashD_{\tilde{\bf k}_1}\langle
J_{{\mu}}(q_{\perp}-k_{\perp1},k_{\perp1};\ell')  \bar{ O}^*_{{\mu}}(\tilde{\bf
k}_1+\tilde{\bf Q}_0)\bar{ O}_{{\mu}}(\tilde{\bf
k}_1) \rangle\cr
 && = \, 2{T\over
LN_\perp}\sum_{{\omega}_n}\sumslashD_{{\bf k}_1}
\big[ J_{{\mu}}(q_{\perp}-k_{\perp1},k_{\perp1};\ell) \cr &&
\ \ \ \ \ \ \ \  \times \ G^0_+({\bf
k}_1 +\tilde{\bf Q}_0,\omega_n)G^0_-({\bf k}_1,\omega_n)
\end{eqnarray}
As a function of $q_\perp$, the  electron and hole of the Peierls loop  
cannot be put simultaneously in the outer energy shell  for all  values of $k_{\perp1}$ in the Brillouin
zone.  Parts of the
$k_{\perp1}$ summation would then refer to $J_\mu$  obtained at previous values of $\ell$. However,
this variation  of
 $J_\mu$  with $\ell$ within the $k_{\perp1}$ interval will be neglected  consistently
with the absence of  dependence on the longitudial momentum for the couplings. After a frequency sum,
one then  finds   
\begin{eqnarray}
F && (q_\perp)=  (\pi v_F)^{-1} {1\over N_\perp}
\sum_{k_{\perp1}}J_{{\mu}}(q_{\perp}-k_{\perp1},k_{\perp1};\ell)\cr && \times \int_{o.s}
{\tanh{\beta\over 2}[E_-
+A(k_{\perp1},q_\perp)] +
\tanh{\beta\over 2} E_-\over 2E_- +A(k_{\perp1},q_\perp)}\, dE_-,
\end{eqnarray}
 where
\begin{eqnarray}
A(k_{\perp1} && ,q_\perp)=  
2t_\perp^*[\cos(k_{\perp1}d_\perp)+\cos(k_{\perp1}d_\perp + q_\perp
d_\perp)  ]
\cr && +2t_{\perp2}^*[\cos(2k_{\perp1}d_\perp) + \cos(2k_{\perp1}d_\perp + 2q_\perp
d_\perp)  ].
\end{eqnarray}
Integrating in  the outer energy shell   yields
\begin{eqnarray}
{1\over2}\,\langle &&(S_{I,2}^P)^2 \rangle_{o.s} =  (\pi v_F) \sum_{\mu }
\sum_{\{ \tilde{\bf
k},\tilde{\bf k}',\tilde{\bf Q} \}^* } {d\ell\over N_\perp}\sum_{k_{\perp1}} 
\cr  && \times \big[ I_P(q_\perp,k_{\perp1},\ell) J_{{\mu}}(q_{\perp}-k_{\perp1},k_{\perp1};\ell)
J_{{\mu}}(q_{\perp}-k_{\perp}',k_{\perp}';\ell)\big] \cr 
&& \ \ \ \ \ \ \times \,{ O}^*_{{\mu}}(\tilde{\bf k} +
\tilde{\bf Q}){ O}_{{\mu}}(\tilde{\bf k}' - \tilde{\bf Q})
\label{Peierls}
\end{eqnarray}
where
\begin{eqnarray}
&& I_P(q_\perp,k_{\perp1},\ell)=  {E_x(\ell)\over 4} \sum_{p=\pm}
\Big[ \ \tanh{\beta\over4} E_x(\ell) + \cr &&
 \tanh{\beta\over2}[E_x(\ell)/2
+pA(k_{\perp1},q_\perp)]\ \Big] 
/[E_x(\ell) +pA(k_{\perp1},q_\perp)].
\end{eqnarray}

\bibliography{articles,Note(theorie)}

\begin{thebibliography}{10}
\bibitem{corrections}
Pair vertex corrections are present in the 1-D regime which yield
  $z_{\mu,(\bar{\mu})} >1$ at $E_x$. Correlations effects are not only present
  in the Peierls channel but also in the interchain Cooper channel -- albeit
  much weaker in amplitude in the latter case (see references
  \cite{Bourbon91,Guay99}).
\end{thebibliography}
\bibliographystyle{prsty}
\end{document}